\documentclass[aip,rsi,reprint]{revtex4-1}
\usepackage[utf8]{inputenc}
\usepackage[english]{babel}

\usepackage{graphicx}
\begin{document}

\title{Arbitrary digital pulse sequence generator with delay-loop timing}

\author{Radim Ho\v{s}\'ak}
\affiliation{Department of Optics, Faculty of Science, Palack\'y University,
             17.\ listopadu 12, 77146 Olomouc, Czech Republic}
\author{Miroslav Je\v zek}
\email{jezek@optics.upol.cz}
\affiliation{Department of Optics, Faculty of Science, Palack\'y University,
             17.\ listopadu 12, 77146 Olomouc, Czech Republic}
\date{\today}

\begin{abstract}

We propose an idea of an electronic multi-channel arbitrary digital sequence
generator with temporal granularity equal to a single clock cycle. We implement
the generator with 32 channels using a low-cost ARM microcontroller and
demonstrate its capability to produce temporal delays ranging from tens of
nanoseconds to hundreds of seconds, with 12~ns timing granularity and linear
scaling of delay with respect to the number of delay loop iterations. The
generator is optionally synchronized with an external clock source to provide
100~ps jitter and overall sequence repeatability within the whole temporal
range. The generator is fully programmable and able to produce digital
sequences of high complexity. The concept of the generator can be implemented
using different microcontrollers and applied for controlling of various
optical, atomic, and nuclear physics measurement setups.

\end{abstract}

\maketitle

%%%%%%%%%%%%%%%%%%%%%%%%%%%%%%%%%%%%%%%%%%%%%%%%%%%%%%%%%%%%%%%%%%%%%%%%%%%
\section{Introduction}

An arbitrary digital pulse sequence generator is a crucial tool for many
instrumentation, automatization, and metrological tasks. It is frequently used
to control and synchronize other devices to form highly complex setups like
magnetic resonance imaging \cite{Brown2014}, trapped ions experiments
\cite{Blatt2008,Monroe2016}, and fast reconfigurable photonic circuits
\cite{OBrien2012} for quantum simulations and quantum communication networks
\cite{Szameit2016,Tanzilli2016}, to name a few applications. A large number of
electronic and optoelectronic building blocks such as switches, digital
attenuators, direct digital synthesizers, digital-to-analog converters,
electro-optic or acousto-optic modulators, and gated solid-state detectors,
which are employed in these setups, require many digital control channels with
precise timing. Commercial solutions typically offer limited flexibility and
impractically large resources per channel. Arbitrary waveform generators (AWGs)
and digital delay generators (DDGs) represent two examples of the frequently
used commercial devices. AWGs allow high degree of sequence control, which goes
far beyond a digital pattern, but are severely limited in sequence length. They
are very inefficient in producing long sparse sequences. DDGs offer picosecond
timing and long delays but are unable to generate complex sequences with many
changes of the output signal.

The lack of flexible digital generators with reasonable timing properties and a
large enough number of output channels leads to the development of custom-built
solutions based on prototyping platforms, such as microcontrollers (MCUs)
\cite{han2007,gas2009,Eyler2011} and field programmable gate arrays (FPGAs)
\cite{sun2013,pru2015,hay2016}. The custom solutions often also offer analog
and radio frequency signals \cite{Eyler2011,pru2015}.
%or extended control functionality \cite{nin2014,cox2016,sau2016}.
MCUs have gained significant computing power, they can be easily programmed,
and their large number of peripherals makes them highly flexible, the downside
of which is that the reaching of full temporal control down to single clock
cycle represents a challenging task. FPGAs offer a single cycle or even
sub-cycle resolution employing a serialization approach \cite{sun2013}.
However, the development on this platform is more complicated compared to the
MCUs, particularly the asynchronous design represents a challenging task.

The design of a custom digital pulse sequence generator, further called
pulsebox, typically utilizes one of three basic approaches: (1)~timer
peripherals, (2)~recalling the sequence from a fast memory, and (3)~communicating
the sequence during a runtime. Using timers to clock delays
between the sequence events \cite{han2007,Eyler2011} offers good temporal
resolution but the corresponding interrupt overhead impacts negatively the
minimum time interval between the events. The fast memory approach allows for
single-clock sequence programming, however, it causes a strict resolution vs.
sequence length trade-off as digital states of all channels for every clock
cycle have to be saved in the memory. This approach is particularly suitable
for short sequences with high temporal resolution \cite{Strachan2007,hay2016}.
Alternatively, the digital states of the pulsebox can be programmed during its
runtime using external communication, which imposes no limitation on the
sequence complexity \cite{sun2013}. However, ongoing communication overhead
limits the minimum time interval between the sequence events and makes such the
solution particularly suitable for long and sparse sequences.

\begin{figure}[t] \centering
  \includegraphics[width=\columnwidth]{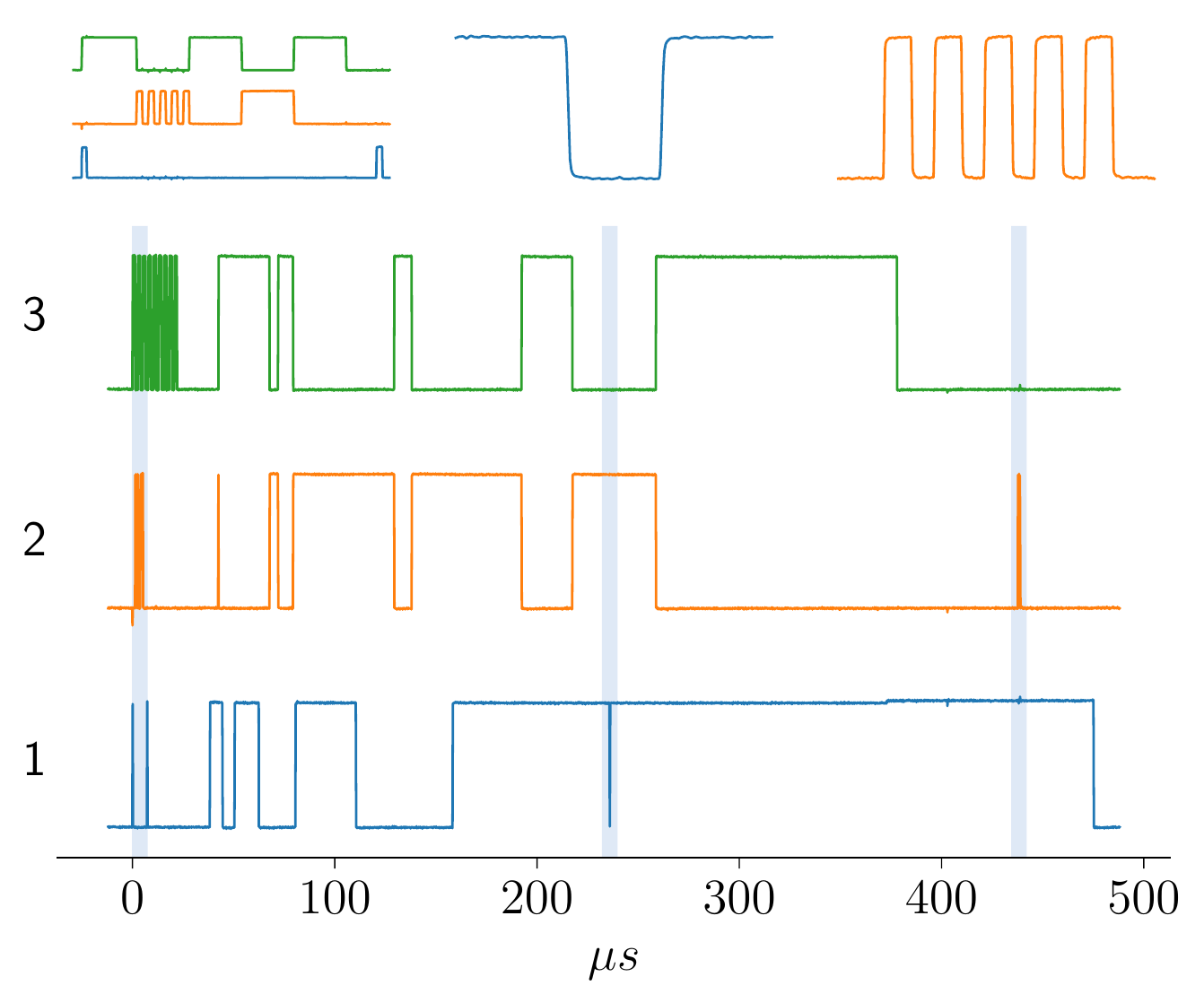}
  \caption{An example of a complex three-channel pulse sequence. The sequence
    pattern consists of both rapid successions of pulses and output changes
    which occur sparsely in time. The highlighted regions near the beginning,
    midpoint, and ending of the sequence are shown in greater detail. The
    shortest structure takes 120~ns, the whole sequence has a duration of
    0{.}5~ms. The depicted waveforms are the actual output of a prototype of
    the presented pulsebox device.}
  \label{fig:compl_seq}
\end{figure}

The presented work focuses on a custom pulsebox device which produces digital
signals on many output channels. The output signals take on the form of
arbitrary pulse sequences, such as the one in Fig. \ref{fig:compl_seq}. The
sequences may contain both rapid successions of short pulses and pulses with
great temporal separation. It is desirable that both these features be
available within the single sequence generated by the pulsebox. We propose here
a concept of this flexible pulsebox, its implementation using an ARM
microcontroller, and a set of characteristics along with methods for their
measurement.

%%%%%%%%%%%%%%%%%%%%%%%%%%%%%%%%%%%%%%%%%%%%%%%%%%%%%%%%%%%%%%%%%%%%%%%%%%%
\section{Design}

A pulse sequence can be understood as a number of digital output changes
occurring at specified times. The pulsebox can reproduce any pulse sequence, as
long as for each output change two pieces of information are known: the desired
digital state of all output channels after the output change, and the time at
which the output change is to occur. Our design relies on a 32bit
microcontroller unit (MCU) whose digital pins are divided into ports of 32 and
individually tied to the pulsebox digital outputs, either directly or via
auxiliary circuitry such as level shifters or buffer amplifiers. The states of
the 32 individual digital pins of each port are governed by a 32-bit register.
Digital pins belonging to the same port are used so that the state of all the
pulsebox output channels can be changed synchronously by manipulating a single
register. The 16 randomly selected channels were characterized during the
calibration stage to demonstrate correct operation. One digital pin of the
MCU is used as an input for an external trigger signal.

To correctly reproduce the pulse sequence, the output changes need to
occur at the specified times. Without artificially introduced delay, two
consecutive output changes will be separated by a time interval corresponding
to the time it takes the MCU to modify the port register. If the output changes
need to be more separated in time, it is necessary to produce an extra delay by
instructing the MCU to stay idle using the NOP assembly instruction
\cite{Pyeatt2016}. This performs a no-operation which takes one clock cycle.
By repeating this instruction in the source code, the distance between two
output changes can be changed by the unit of one clock cycle. This is the best
granularity possible for an MCU-based pulsebox.

Using the NOP instruction yields very good granularity, but is not desirable
for long delays, as the MCU program memory would fill up very quickly. Instead,
the NOP instruction is looped over inside a delay loop. The loop can be
repeated almost infinitely, making a very large temporal distance between
consecutive output changes possible. However, the loop consists of multiple
instructions. This means that the granularity is worse than when using NOP
instructions exclusively.

The delay loop consists of two distinct parts---the setup and the looped
section. The setup specifies the number of loop iterations and its processing
by the MCU takes a constant amount of time. The looped section is repeated a
given amount of times and is responsible for most of the delay achieved via the
delay loop. The time spent in the delay loop setup gives a lower bound on the
length of the shortest pulse achievable using this method, and the length of
the looped section determines the granularity with which we can vary the delay
lengths. The loop has thus been designed to minimize the number of instructions
in both the setup and the looped section, leading us to implement it in
assembly language.

To access the entire range of achievable temporal distances between consecutive
output changes, and still benefit from the one-cycle granularity, both the NOP
approach and the delay loop approach must be combined. The only drawback here
comes from the way the MCU predicts which instructions will be performed next.
When conditional branches occur, which is the case for the delay loops, the
program execution might stall up to three clock cycles before the MCU discovers
which instructions follow. This could lead to an up to three-clock deviation in
the output change timing \cite{ArmM3}. Fortunately, this is a systematic
deviation and not a jitter, as the timing stays constant throughout sequence
repetitions.  This effect can be eliminated by a detailed calibration that
would provide feedback for the sequence input.

The final part of the design is the software package for user-friendly
reconfiguration of the device. The software gathers user-provided data about
the shape of the sequence, generates a source code file with instructions for
the pulsebox to reproduce the sequence, compiles it, and uploads the resulting
binary file to the MCU. The software is able to extract the sequence shape data
from different user-provided information, such as the positions and lengths of
individual pulses. The source code is created by gluing together instructions
for changes of states of the MCU digital pins with delay loops that provide the
MCU with idle time between output changes.

Our physical realization of the concept mentioned above relies on the Arduino
Due development board based on the Atmel SAM3X8E microcontroller \cite{SAM3X8E}.
Arduino Due has been previously used for custom-built control solutions in several
advanced applications including optical tweezers, preparation and measurement
of non-classical states of light, and quantum memory \cite{nin2014,cox2016,sau2016}.
Implementing the pulsebox on this low-cost development board demonstrates
the feasibility of the delay-loop concept and, also, further extends the
applicability of similar MCUs for advanced analog as well as digital control.

%%%%%%%%%%%%%%%%%%%%%%%%%%%%%%%%%%%%%%%%%%%%%%%%%%%%%%%%%%%%%%%%%%%%%%%%%%%
\section{Pulsebox performance}

To characterize the performance and capabilities of the device, we have chosen
these figures of merit: (1)~the minimum and (2)~maximum time interval between
output changes, (3)~timing granularity, (4)~linearity, (5)~run-to-run
uncertainty, (6)~inter-channel simultaneity, (7)~trigger latency, and
(8)~maximum sequence complexity.

\begin{figure}[t] \centering
  % \includesvg[width=\columnwidth]{pics/noext_clk_cal_line_v4}
  % \includegraphics[width=\columnwidth]{pics/noext_clk_cal_line_v4}
  \includegraphics[width=\columnwidth]{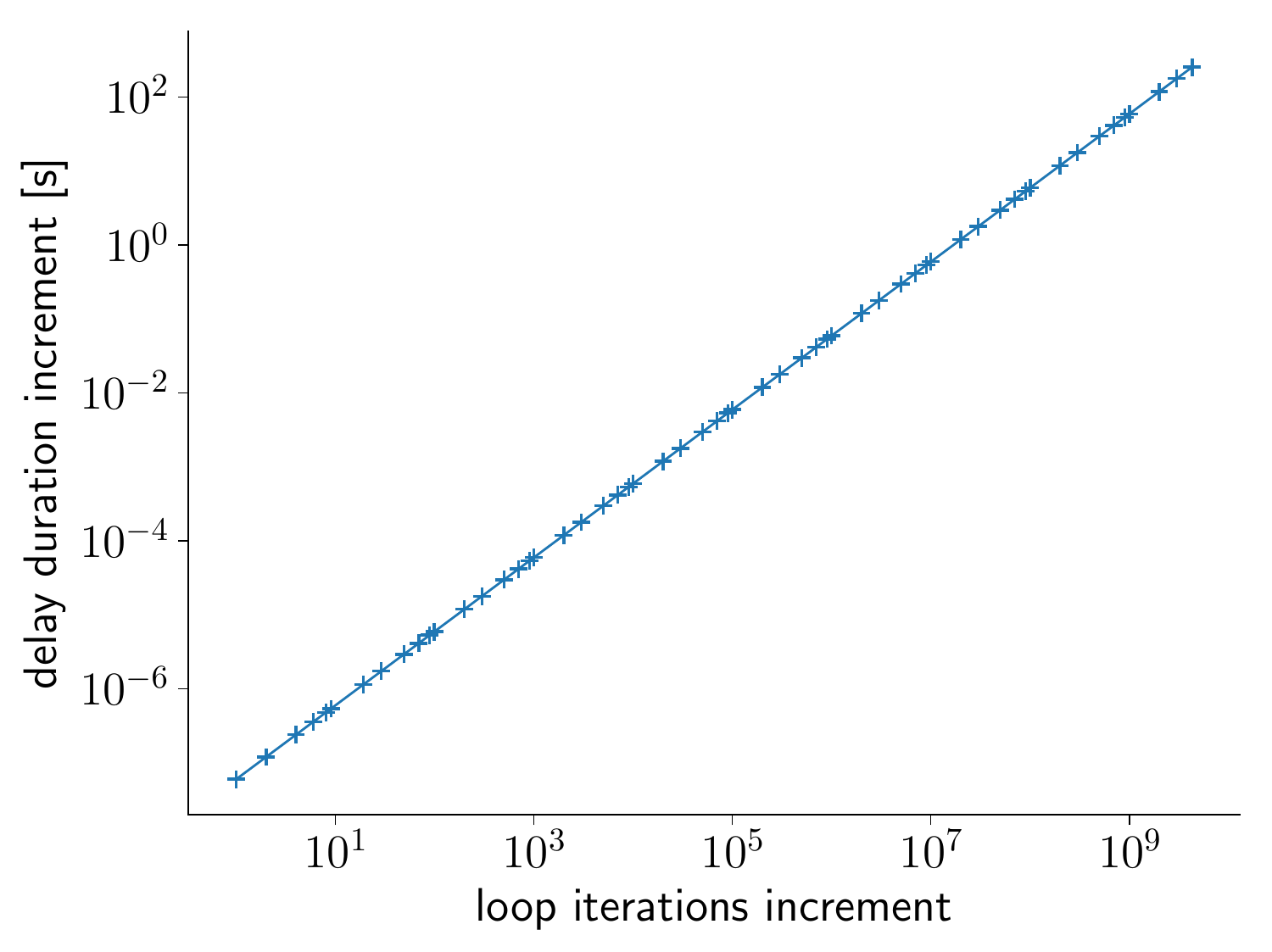}
  \caption{The delay duration increment is shown as a
    function of the increment in a number of delay-loop iterations. The
    calibration data are fitted to a linear function.
    %$y = a \times x + b$, with $a = 60\ \mathrm{ns}$, $b = -212\ \mathrm{ps}$.
    See text for more details.
    }
  \label{fig:noext_clk_calibration_line}
\end{figure}

\begin{figure}[t] \centering
  % \includesvg[width=\columnwidth]{pics/noext_clk_cal_spandev_v4}
  % \includegraphics[width=\columnwidth]{pics/noext_clk_cal_spandev_v4}
  \includegraphics[width=\columnwidth]{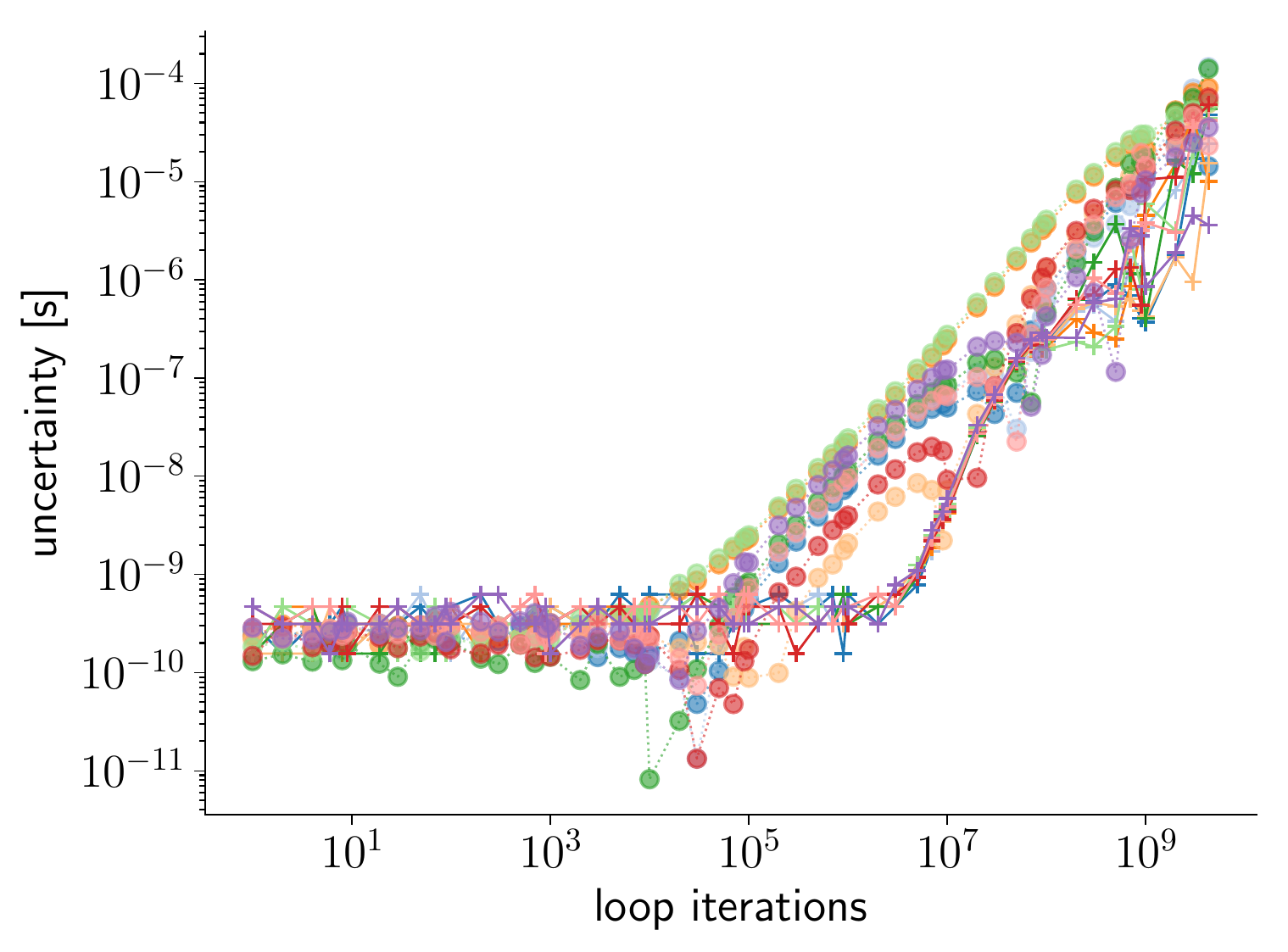}
  \caption{The delay duration span (cross marks, solid
    lines) and the absolute value of the mismatch between the data and the
    calibration line fit (circle marks, dotted lines) are shown for each
    calibration run. An increase in both the delay span and the fit mismatch,
    caused by the drift and/or instability of the internal clock, is seen
    after $10^5$ iterations.}
  \label{fig:noext_clk_calibration_span_and_dfit}
\end{figure}

To measure \emph{the minimum time interval between output changes}, an output
change, and then immediately another one, were made on a single channel,
resulting in a single pulse. The minimum pulse length was then the time
interval between the 50\% crossings of the rising and falling edge of the
pulse, which we measured using an oscilloscope. For our device, the result was
24~ns, which corresponds to two clock cycles of the MCU. Using the delay loop
approach, the minimum pulse length was achieved using a single iteration loop
and was measured to be 144~ns, or 12 clock cycles.

The device achieves the ultimate \emph{granularity} using the NOP approach. By
incrementally adding NOP instructions between two output changes, we observed
prolonging of the pulse in steps of 12~ns, or one clock cycle. Additionally,
it is of interest to know the granularity of the delay loop approach. Instead
of inferring this information in the same manner as in the case of the repeated
NOP instructions, we perform a complete calibration of the dependence of delay
duration on the number of delay loops.

The delay-loop calibration relies on measurements of the delay between two
consecutive output changes for different numbers of delay loops, starting from
one iteration and going up to the maximum number. The delay was measured using
a time-to-digital converter (TDC, UQDevices) sensitive to rising edges only, so
a two-pulse sequence was used for the measurements, the delay loop being placed
between the pulses. To get the delay between the falling edge of the first
pulse and the rising edge of the second one, and more specifically, the
prolonging of the said delay with various increases of the number of delay-loop
iterations, we took the TDC result for one-iteration delay loop as a reference,
and subtracted it from the results for the other numbers of iterations. The
final result yields the calibration curve showing linear dependence for delay
loops, which tells us by what amount is the delay between two output changes
prolonged if the number of iterations of the corresponding delay loop is
increased from one to a certain amount, see
Fig.~\ref{fig:noext_clk_calibration_line}.

The whole calibration procedure was repeated 9 times to verify the stability
and the linearity of the delay-loop approach. For each loop-iteration increment
we evaluated a min-max span of the corresponding delay duration across all the
measurements, which characterizes the \emph{run-to-run uncertainty} of the
given delay, see Fig.~\ref{fig:noext_clk_calibration_span_and_dfit}. In
our case, the uncertainty is of the order of 100~ps for delays up to 100~ms and
increases rapidly for larger delays.

To verify the perfect \emph{linear dependence}, the acquired data were fitted
to a linear function $y = a \times x$. From the slope of the linear fit, we
estimated the delay-loop timing granularity to be 60~ns, or 5 clock cycles.
A mismatch between individual calibrations and the linear fit shows similar
features as the run-to-run uncertainty, see
Fig.~\ref{fig:noext_clk_calibration_span_and_dfit}. We will show later that the
increase of the uncertainty as well as of the mismatch is caused by internal
clock drift and can be entirely removed using external clock source.

\emph{The maximum time interval between output changes} was obtained from the
calibration data for the delay loop with the highest possible number of
iterations, with the resulting value of 255{.}7~s.

The \emph{inter-channel simultaneity} of output changes has also been studied.
When all pulsebox channels were set to perform the change from a low voltage
level to a high voltage level, the time difference between the rising edges at
50\% signal level was found to be 2~ns at maximum between the slowest and the
fastest channel. This value represents an upper bound on the simultaneity
including all the inter-channel delays originated in the pulsebox, cables, and
connectors.

\emph{The trigger latency} has been measured by running a single-pulse sequence
in the triggered mode of operation. A signal from an external generator is fed
to the pulsebox. A copy of the signal is used to trigger an oscilloscope. The
trigger latency is the time interval between the 50\% crossings of the trigger
signal and the first pulse of the pulsebox output. With the device in its
current form, the trigger latency is around 900~ns, with roughly 60~ns
peak-to-peak jitter. These values are more than sufficient taking into account
that the pulsebox is typically triggered by external events with periods much
larger than 1~$\mu s$, for example to synchronize the pulsebox with the power
line frequency or with data acquisition. Once triggered, the output changes at
all channels within a single sequence are generated with negligible sub-ns
run-to-run uncertainty and used to control all the circuitry and devices in the
setup.

\emph{The maximum sequence complexity} is characterized in terms of the maximum
number of output changes and delay loops that can occur in a sequence. With
rising complexity, the MCU program memory gradually fills up, until its
capacity is exceeded. The maximum complexity was estimated using a model pulse
sequence which uses a delay loop between each two consecutive output changes.
For our current implementation the most complex sequence of this kind that we
were able to generate consisted of 10670 output changes and 10669 delay loops.

%%%%%%%%%%%%%%%%%%%%%%%%%%%%%%%%%%%%%%%%%%%%%%%%%%%%%%%%%%%%%%%%%%%%%%%%%%%
\section{Operation with external clock source}

By default, the Arduino Due development board runs with 84~MHz master clock
frequency, which is obtained from a 12~MHz crystal oscillator using clock
dividers and multipliers. Replacing this crystal with an external clock makes
it possible to synchronize the operation of the pulsebox with other devices
used in the experimental setup and eliminate the drift of the internal clock.
Instead of the common 10~MHz external clock we used a 12~MHz one to facilitate
the comparison of the pulsebox timing properties with the internal clock.

\begin{figure}[t] \centering
  % \includesvg[width=\columnwidth]{pics/ext_clk_cal_spandev_v4}
  % \includegraphics[width=\columnwidth]{pics/ext_clk_cal_spandev_v4}
  \includegraphics[width=\columnwidth]{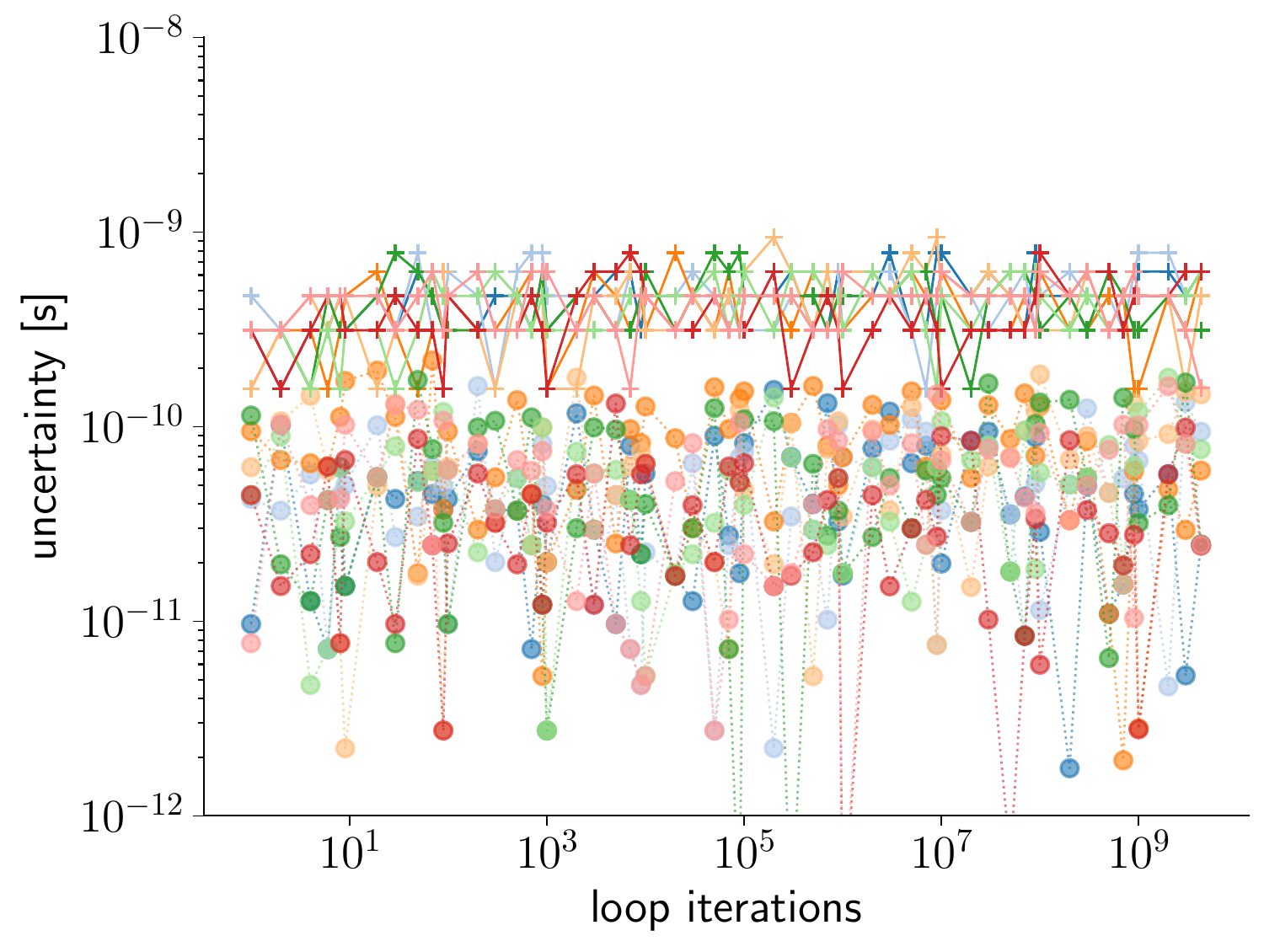}
  \caption{Calibration results for pulsebox operating with an external clock
    source. The delay duration span (cross marks, solid lines) and fit mismatch
    (circle marks, dotted lines) are presented, demonstrating perfectly stable
    operation with sub-nanosecond jitter.}
  \label{fig:ext_clk_calibration}
\end{figure}

The calibration procedure has been repeated 8 times for the pulsebox with the
external clock source, and indeed it shows a sub-nanosecond run-to-run
uncertainty over the entire range of delays, see Fig.
\ref{fig:ext_clk_calibration}. The fit mismatch was also reduced below one
nanosecond for all the acquired measurements and all possible delays of the
pulsebox.

% \begin{figure}[t] \centering
%   \includegraphics[width=\columnwidth]{pics/ext_clk_calibration_line}
%   \caption{Calibration curve (external clock source).}
%   \label{fig:ext_clk_calibration_line}
% \end{figure}

%%%%%%%%%%%%%%%%%%%%%%%%%%%%%%%%%%%%%%%%%%%%%%%%%%%%%%%%%%%%%%%%%%%%%%%%%%%
\section{Possible extensions}

The functionality of the device can be further extended and some of the
characteristics can be improved.
As for the number of digital output channels, it is possible to go beyond 32.
However, this means that more digital pin ports must be used and that multiple
registers will sometimes have to be modified in order to perform an output
change. This automatically leads to worse inter-channel output change
simultaneity, as multiple corresponding registers cannot be changed using a
single instruction. This can, however, be compensated in post-processing using
a digital delay. The upper bound on the distance between two consecutive output
changes can be overcome by either performing multiple delay loops or by nesting
a delay loop inside another loop. The nested loop approach would yield a
substantial increase of the range of possible delays.

Using an external clock source opens up the possibility of overclocked device
operation. Higher clock rate allows for shorter pulses and lower granularity.
The datasheet of the used MCU specifies the external oscillator frequency to be
in the range of 3--20~MHz. This frequency is, by default, effectively
multiplied by a factor of 7. Overclocked, and even underclocked, operation has
been achieved in our tests. However, we have observed that uploading code to
the device fails for some frequencies of the external clock source. This can be
avoided by changing the clock frequency to the 12~MHz default for the upload
and manipulating the clock frequency only afterwards. Furthermore, the
operation stopped if the master clock frequency was too high. The master clock
frequency can be further manipulated by changing the quotients of the built-in
frequency dividers and multipliers, which will be subject to further tests.
Alternatively, a faster-clock supporting MCU can be used to reach even better
granularity.

The presented pulsebox generates unipolar signals. Bipolar operation can
be attained using a two-way combiner to merge two unipolar channels
into a single bipolar one, which was demonstrated for pulse generators
by Strachan et al. and Haylock et al.\cite{Strachan2007,hay2016}.

Finally, some of the channels could be used to feed programmable
attenuators/amplifiers, direct digital synthesizer, or digital-to-analog
converters. This would enable radio-frequency or arbitrary analog outputs. The
performance of these outputs, such as the minimum distance between two signal
changes, would depend on the parameters of the employed extension device. We
tested interfacing the pulsebox to programmable digital attenuator operating
from 9~kHz to 6~GHz (Mini Circuits ZX76-31R75PP+).
% with dynamical range of 32~dB and switching speed of 300~ns.
Using direct parallel programming, we can command the attenuator to set a new
attenuation level by sending 7bit digital word within one clock cycle of our
pulsebox \cite{Straka2017}.

%%%%%%%%%%%%%%%%%%%%%%%%%%%%%%%%%%%%%%%%%%%%%%%%%%%%%%%%%%%%%%%%%%%%%%%%%%%
\section{Conclusion}

We have presented a general-purpose digital pulsebox capable of producing
complex sequences of pulses both in rapid succession and with great temporal
separation, which sets it apart from devices built for one specific
application. We have implemented the pulsebox using a low-cost ARM
microcontroller and demonstrated its wide functionality, considerable
precision, and sub-nanosecond timing jitter. The pulsebox timing granularity is
12~ns, or one clock cycle, achieved by using delay loops together with
single-cycle no-operation instructions. The presented 24~ns (2 clock cycles)
minimum interval between output changes is a significant improvement over
many-cycle intervals presented by many previous works. The employed delay-loop
approach allows for extensive delays up to 255 s between the output changes
while preserves the high temporal resolution and prevents excessive memory
usage. The information about the shape of the sequence is communicated to the
pulsebox by storing a specific program code in the memory. This approach,
combined with the program memory capacity available on the MCU employed, yields
the maximum sequence complexity of more than 20{,}000 output changes and delays.

Apart from designing the pulsebox itself, a set of characteristics and the
methods of their measurement, applicable to a broad spectrum of similar
devices, has been devised. We believe that these basic characteristics will
facilitate characterization and comparison of future multi-channel digital
pulse sequence generators.\\

\acknowledgments

This work was supported by the Czech Science Foundation (project 17-26143S).
RH also acknowledges the support by the Palacky University (project
IGA-PrF-2017-008). We would like to thank M.~Dudka, I.~Straka, P.~Ob\v{s}il,
and L.~Slodi\v{c}ka for many fruitful discussions on the pulsebox concept and
applications.

% %\bibliographystyle{plain}
% \bibliographystyle{aipnum4-1}
% %\setcitestyle{numbers,square}
% \bibliography{pb_article}
 
%

\end{document}